# Entangled Space-Time

**Paola Zizzi**
*Department of Brain and Behavioural Sciences, University of Pavia,
Piazza Botta, 11, 27100 Pavia, Italy*

**Abstract**
We illustrate the entanglement mechanism of quantum space-time itself.
We consider a discrete, quantum version of de Sitter Universe with a Planck time-foliation, to which is applied the quantum version of the holographic principle (a Planckian pixel encodes one qubit rather than a bit). This results in a quantum network, where the time steps label the nodes. The quantum fluctuations of the vacuum are the connecting links of the quantum network, while the total number of pixels (qubits) of a spatial slice are the outgoing links from a node n.
At each node n there is a couple of quantum gates, the Hadamard gate (H) and the controlled-not (CNOT) gate, plus a projector P. The Hadamard gate transforms virtual states (bits) into qubits, the projector P measures a qubit at the antecedent node, giving rise to a new bit, and the CNOT gate entangles a qubit at node n with the new bit at node n-1.
We show that the above quantum-computational interpretation of space-time entanglement has a geometrical counterpart. In fact, the quantum fluctuations of the metric on slice n are such that a tiny wormhole will connect one Planckian pixel of slice n with one of slice n-1. By the quantum holographic principle, such a geometrical connection is space-time entanglement.

# 1. Introduction

Entanglement, a very particular property of the quantum world, is a quantum correlation that does not have an analogue in the classical world. The building blocks of quantum entanglement are the qubits (we recall that a logical qubit is the unit of quantum information, the quantum analogue of the classical bit, that is, a quantum superposition of bits 0 and 1).

An entangled bipartite quantum state, in which the two parts are the qubits A and B, is not separable, that is, it can not be written as a tensor product $A \otimes B$.

The physical realization of a logical qubit is for example a spin ½ particle (electron) or single atoms or ions (with two internal electronic states) or single polarized photons.

Since now entanglement has been demonstrated experimentally with optical photons, neutrinos, and electrons.

But, in the context of quantum gravity, the following question arises: can entanglement also concern quantum space-time itself?

The answer depends on the context, but the important task is to make the very concept of space-time entanglement as clear as possible. To speak of an entangled space-time means first of all to consider a discrete space-time, possibly given in integer multiples of Planck units of time and length. This request is necessary to associate to each Planckian pixel, at a given time step, a logical qubit B that could be entangled with another qubit A at the precedent time step. As we already said, an entangled bipartite quantum state is not separable. This means that the two parties lose their identity and behave as a whole. In the case of space-time entanglement, then, two pixels of area at different time steps would behave as a whole.

The non-separable character of entangled space-time would lead us to review many of our established beliefs, such as locality, the arrow of time and causality.

Note that the simple fact that pixels encode qubits is a necessary but not sufficient condition for spatio-temporal entanglement. An entanglement mechanism is required, and the only operation that can provide it is the transformation carried out by the quantum port CNOT (Controlled Not). The CNOT gate uses a qubit B as a control and a bit $0_A$ (or $1_A$) as a target and returns a maximum entangled state (a Bell state).

A question would then arise: where that bit appears from, as all pixels encode qubits, rather than classical bits? The answer is that bit $0_A$ (or $1_A$) is obtained by measuring the qubit A. Then, a two-dimensional projector is required as well. Another question could be the following: why should we expect that space-time is entangled? The answer is that, in the case of discrete space-time, entanglement is what "glues" together spatial slices occurring at different time steps. Finally, one might wonder which is the physical mechanism of entanglement, that is formally simulated by the operations of the CNOT gate and the projector. As we will see in this paper, such a mechanism is lead by the quantum fluctuations of the vacuum, together with the quantum fluctuations of the metric.

We believe, then, that at the fundamental level of the Planck scale the answer to the question whether space-time itself can be entangled is affirmative, and in this paper we will give the motivations of our belief. But first, we would like to make the following remark.

From the fact that quantum entanglement is a quantum correlation which has not a classical analogue, it follows that the appearance of entanglement in a theory under study ensures that such a theory is a quantum theory. Then, if a space-time theory manifests an entangled structure, we would be sure that it is a quantum space-time theory, a candidate for quantum gravity.

The motivations that we will give in this paper are based on various concepts and results already illustrated in previous works, among which the discrete, quantum version of the empty space-time of the de Sitter universe [1], its logical quantum-computational realization in terms of a Quantum Growing Network (QGN) [2] and the quantum extension [3] of the holographic principle [4]. In the context of Loop Quantum Gravity (LQG) [5] the formalism of "spin networks" leads to the very important result of discreteness of area and volume [6]. The application of the quantum holographic principle (QHP) [3] to the formalism of spin networks in LQG, lead to Computational Loop Quantum Gravity (CLQG) [7].

In this work, we will make a change to the QGN, by including an internal observer who, standing on the n[th] horizon of the de Sitter's discrete universe, observes the (n-1)[th] horizon by using a photon with the appropriate energy. The presence of the observer is equivalent to add a projector to node n, where there was already a Hadamard quantum gate. The apparent loss of the quantum information due to the measurement is restored by the quantum gate CNOT also added to node n, which entangles a qubit of node n with one of node n-1, by using the bit, obtained from the measurement, as target. This new quantum network will be called OQGN, where "O" stands for "Observer". It may seem that the introduction of the CNOT quantum gate to preserve quantum information through entanglement is done by hand. Instead, it is only the logical aspect of what happens physically. In fact, the energy of the vacuum is shared between the energy of the observational photon (OP) and the cosmological constant. Since the latter is given in terms of quantum information, this energy balance ensures the conservation of quantum information. In turn, the quantum information required for the balance is given by the entanglement entropy of a Bell state. In logical terms, this is just given by the action of the quantum gate CNOT.

This modification of the QGN, which leads to entanglement of a qubit of node n with a qubit of node n-1, can be also seen in a geometrical way, by using Wheeler's quantum foam [8] in terms of the quantum fluctuations of the metric, and the QHP.

We will show that the discrete quantum fluctuations of the metric on the n[th] spatial slice have a quantum-gravitational energy with a discrete spectrum. This quantum-gravitational energy is "borrowed" from the energy of the quantum fluctuations of the vacuum, that is, from the cosmological constant. And the energy balance guarantees the conservation of quantum information. For a certain expression of the quantum-gravitational energy, equal to that of the OP, this equilibrium leads to entanglement.

What get entangled in this geometric version? Space-time itself. In fact, a Planckian pixel of slice n gets entangled with one of slice n-1, because of the QHP.

The entangled pixels are in fact identified with each other, through virtual wormholes, which are the maximum quantum fluctuations of the metric at the Planck scale.

Thus, the two equivalent views of space-time entanglement discussed in this paper seem to fit very well with the ER-EPR conjecture [9].

The paper is organized as follows.

In Sect. 2, we give a short review of the original QGN.

In Sect. 3, we illustrate the new OQGN, achieve entanglement in a logical quantum-computational way, and discuss the physical implications.

In Sect. 4, we give the geometrical interpretation (in terms of the discrete quantum fluctuations of the metric) of the results obtained in Sect. 3.

In Sect. 5, we illustrate space-time entanglement in terms of virtual wormholes.

Sect. 6 is devoted to the conclusions.

## 2. The QGN: A brief review

The QGN was based on a quantum, discrete version [1] of the de Sitter Universe: space and time are both discrete. Each time step $t_n$ is an integer multiple of the Planck time, and the spatial dimension at time $t_n$ is an integer multiple of the Planck length:

$$t_n = (n+1)t_P \qquad L_n = (n+1)l_P \qquad (2.1)$$

where $t_P \cong 10^{-43}$ sec is the Planck time, and $l_P \cong 10^{-35} m$ is the Planck length.

## 2.1 The basics of the QGN

At each time step, the de Sitter horizon has a discrete area $A_n = (n+1)^2 l_P^2$ given in terms of $(n+1)^2$ pixels (a pixel is one unit of Planck area, $l_P^2$ ). The discrete entropy satisfies the Bekenstein bound [10].

The quantum fluctuations of the metric $\Delta g$ are discrete, and given by the relation:

$$\Delta(g)_n = \frac{\Delta g_n}{g_n} = \frac{1}{n+1} \qquad (n=0,1,2,3...). \tag{2.2}$$

The cosmological constant is also quantized, and given in terms of quantum information:

$$\Lambda_n = \frac{1}{I_n l_P^2} \tag{2.3}$$

where $l_P$ is the Planck length, $I_n$ is the quantum information at time $t_n$:

$$I_n \equiv N = (n+1)^2 \tag{2.4}$$

and N is the total number of qubits.

The interesting aspect of this discrete model, is that it is a quantum-computational space-time, as it does support the QHP. The latter is the quantum version of the Holographic Principle, where instead of interpreting each pixel of area as a classical bit of information, one interprets it as a qubit. This is possible if the horizons' surfaces are pierced by edges of spin networks [6] labelled by the spin-1/2 representation of SU(2) in the superposed state of spin "on" and spin "down".

This discrete de Sitter universe can be interpreted as a special kind of quantum memory register, where the evolution time is discrete. Thus, the usual quantum logic gates are replaced by discrete unitary evolution operators which connect Hilbert spaces of different dimensionality.

In [2], the (discrete) early inflationary universe [1] was described as a quantum growing network (QGN). The speed up of growth of the network (inflation) is due to the presence of virtual qubits in the vacuum state of the quantum memory register. Virtual quantum information is created by quantum vacuum fluctuations, because of the inverse relation between the quantized cosmological constant $\Lambda_n$ and quantum information $I_n$ given in Eq. (2.3).

At each time step $t_n$, the increase of quantum information from slice n to slice n+1 is given by:

$$\Delta I_{n+1,n} = I_{n+1} - I_n = 2n+3. \tag{2.5}$$

By Eq. (2.3), the increase $\Delta I_{n+1,n}$ of quantum information from slice n to slice n+1, is related to the variation $\Delta \Lambda_{n+1,n}$ of the cosmological constant, that is, to quantum fluctuations of the vacuum. Then, the $2n+3$ units of quantum information in Eq.(2.5) are not available at time $t_n$, as they are virtual states (VS) of the vacuum. They will be transformed, by a quantum logic gate $U_n$ into $2n'+1$ (with $n'=n+1$) available quantum information (QI) at the time step $t_{n+1}$:

$$2n+3 \text{ VS at time } t_n \xrightarrow{U_n} 2n'+1 \text{ QI at time } t_{n'} = t_{n+1} \tag{2.6}$$

where:

$$U_n = \prod_{j=1}^{v=2n+3} Had(j) \tag{2.7}$$

and *Had(j)* is the Hadamard gate:

$$Had = \frac{1}{\sqrt{2}} \begin{pmatrix} 1 & 1 \\ 1 & -1 \end{pmatrix} \tag{2.8}$$

operating on bit j.

Then, the $2n+3$ virtual states at time $t_n$ are transformed into $2n'+1$ qubits at time $t_{n'}$.

For example, at time $t_0$ there are three virtual states, which will be transformed into three qubits at time $t_1$; at time $t_1$ there are five virtual states which will be transformed into five qubits at time $t_2$, and so on.
This results in a quantum growing network, where the nodes are the quantum logic gates, the connecting links are the virtual states, and the outgoing links are the qubits.
At each time step, the total number of qubits, that is, of outgoing links is $N = (n+1)^2$.
As it was shown in [2], the QGN saturates the quantum limits to computation [11].
The rules of the QGN are resumed below.

## 2.2 The rules of the QGN

At the starting time (the unphysical time $t_{-1} = 0$), there is one node, call it **-1**. At each time step $t_n$, a new node is added, which links to the youngest and the oldest nodes, and also carries 2n+1 outgoing links. Thus, at the Planck time $t_0 = t_P$, the new node **0** is added, which links to node **-1** and carries one outgoing link. At time $t_1 = 2t_P$, the new node **1** is added, which links to nodes **-1** and **0,** and carries three outgoing links, and so on. In general, at time $t_n$, there are:
1) $n+2$ nodes but only n+1 of them are active, in the sense that they have outgoing links (node **-1** has no outgoing links).
2) $N = (n+1)^2$ outgoing links coming out from n+1 active nodes.
3) 2n+1 links connecting pairs of nodes.
4) n loops.
In summary:
Node n is a quantum logic gate $U_n$.
The 2n+1 links outgoing from node n are qubits.
The 2n+1 connecting links from node -1 to node n are virtual states.
See **Fig.1**.

## 3. The OQGN (Observed QGN)

To start, it should be said that, formally, the OQGN is obtained from the QGN by associating to each node *n*, at time $t_n$, an hypothetical internal observer $O_{n,n-1}$, which acts on node n-1.
We shall illustrate the simplest case $O_{1,0}$ that is, the observer being at node 1 at time $t_1$, where there are three outgoing links representing three cat states, let us call them $|Q\rangle_B$, $|Q\rangle_C$, $|Q\rangle_D$, and acting on node 0, where there is one outgoing link representing one cat state $|Q\rangle_A$ of the 2-dimensional Hilbert space $C^2$:

$$|Q\rangle_A = \frac{1}{\sqrt{2}}(|0\rangle + |1\rangle)_A .$$

### 3.1. The local action of the internal observer

The observer $O_{1,0}$ measures the qubit $Q_A$ by projecting it, for example on the (normalized) state $|0\rangle_A$ (or $|1\rangle_A$), that is:

$$P_0 Q_A = |0\rangle_A \quad \text{(or } P_1 Q_A = |1\rangle_A \text{)} \quad \text{(both with probability } \tfrac{1}{2}\text{)} \tag{3.1}$$

where $P_0$ and $P_1$ are the two projectors of the 2-dimensional Hilbert space $C_A^2$:

$$P_0 = \begin{pmatrix} 1 & 0 \\ 0 & 0 \end{pmatrix}, \quad P_1 = \begin{pmatrix} 0 & 0 \\ 0 & 1 \end{pmatrix}. \tag{3.2}$$

In what follows, we will consider the case with projector $P_0$. In this case, then, we are left with the bit $|0\rangle_A$ on node $n = 0$.

Now, let us consider the CNOT gate at node $n = 1$. At node 1 there are three cat states, each one of them being a possible control for the CNOT gate.

Let us take, for example, the cat-state $|Q\rangle_B$ to be the control at node $n = 1$, and the bit $|0\rangle_A$ at node $n = 0$ will be the target.

The CNOT flips the target when the control is $|1\rangle$ and leaves the target unchanged when the control is $|0\rangle$:

$$|1\rangle_A \otimes |0\rangle_B \xrightarrow{CNOT} |1\rangle_A \otimes |1\rangle_B \qquad (3.3)$$
$$|0\rangle_A \otimes |0\rangle_B \xrightarrow{CNOT} |0\rangle_A \otimes |0\rangle_B.$$

Eventually, the CNOT will entangle the cat-state $|Q\rangle_B$ at node 1 with the bit $|0\rangle_A$ left at node 0 after the measurement, giving rise to the Bell state:

$$\frac{1}{\sqrt{2}}\left(|0\rangle_A \otimes |0\rangle_B + |1\rangle_A \otimes |1\rangle_B\right).$$
(3.4)

This describes the maximal entanglement of a discrete spatial slice at time $t_{n-1}$ with the one at time $t_n$.

**3.2 Entanglement of outgoing links**

Let us consider, for example, node 1 at time $t_1$, which can be considered as the "present" instant. One of the three outgoing links of node 1 is entangled with the outgoing link of node 0 at time $t_0$. In this way, the "present" $t_1$ shares one unit of quantum information with its most recent past $t_0$, thus incrementing the "memory" of the past. It would seem at this point, that node 1 is left with two outgoing links, but one of them gets entangled with one of the five outgoing links of node 2 at time $t_2$, which is the immediate "future" of $t_1$. Node 1 receives one unit of quantum information from its entangled immediate future. Then, node 1 has still three outgoing links, but two of them are entangled, one with the most recent past and one with the immediate future. There is left only one non entangled outgoing link, as it was in its most recent past (node 0) in absence of observation, as in the original QGN. In the same way, node 2 would receive a unit of quantum memory from its immediate future (node 3). Eventually, node 2 will have still five outgoing links: two of them entangled, one with its most recent past and one with its immediate future, and only three not entangled outgoing links, as it was for its most recent past (node 1) in absence of observation, and so on. The only node that cannot share quantum information with the past is node 0, because its most recent past (node -1) is unphysical.

**3.3 The rules of the OQGN**

The rules of the OQGN are resumed below.
In general, at time $t_n$ there are:

1) 2n+1 nodes, but only 2n of them are active (in the sense that they have outgoing links) in fact node -1 is unphysical, and has no outgoing links.

Each node represents two quantum gates (a Hadamard gate H and a CNOT gate) and a projector P. At time $t_n$ the Hadamard gate H transforms a virtual state $|v\rangle_A$ into a qubit $|Q\rangle_A$ (as in the original QGN), a projector $P_0$ (or $P_1$) projects the qubit $|Q\rangle_A$ into the bit $|0\rangle_A$ (or $|1\rangle_A$) and a CNOT gate entangles one outgoing link at time $t_n$, with one at time $t_{n-1}$, by using the qubit $|Q\rangle_B$ at node n as a control, and the bit $|0\rangle_A$ (or $|1\rangle_A$) at node n-1 as a target.

Then, the sequence of those quantum-classical-quantum operations at each node is:

H------→P-----→CNOT  (3.5)

with:

$$H : |v\rangle_A \to \frac{1}{\sqrt{2}}(|0\rangle + |1\rangle)_A$$

$$P_0 : \frac{1}{\sqrt{2}}(|0\rangle + |1\rangle)_A \to |0\rangle_A \quad (3.6)$$

$$CNOT : \begin{cases} \frac{1}{\sqrt{2}}(|0\rangle + |1\rangle)_B \\ |0\rangle_A \end{cases} \to \frac{1}{\sqrt{2}}(|0\rangle_A|0\rangle_B + |1\rangle_A|1\rangle_B).$$

At time $t_0$, there is only the Hadamard gate H but there are neither the projector P nor the CNOT gate, because node -1 is unphysical.

2) At each node n there are $2n-1$ outgoing links not entangled, and two entangled outgoing links, one entangled with one outgoing link of node n-1, and another entangled with one outgoing link of node n+1.

3) $3n+1$ links connecting pairs of nodes. Note that in the original QGN there were only 2n+1 connecting links. Due to entanglement, n extra connecting links are added.

4) $n$ loops.

See **Fig. 2.**

### 3.4 Entangled time

In this discrete quantum model of empty space-time, the "flow" of time seems to appear in the presence of entanglement.

However, what really arises is a flow of quantum information. This can be seen as a "flow" of time due to the sharing of quantum information between future and past nodes. In this way the past acquires its quantum memory from the future: it should be better interpreted as an "inverse arrow of time".

Without an internal observer, this discrete quantum space-time would seem static although the Planck clock definitely touches.

The presence of a fictitious internal observer actually describes a quantum program, which on a first Hadamard gate network and a projective measurement of a qubit uses an additional network of quantum logic gates, i.e., CNOT ports.

The internal observer, observing the Planck scale, sees a quantum space-time structure, which is formally the same early quantum universe illustrated in the original QGN. However, while in the QGN there is no flow of quantum information, in OQGN there is. This is due to the use by the internal observer of CNOT gates and projectors. Any metaphysical external observer would not be able to make measurements (use of projectors) or remain entangled with anything (use of the CNOT gate). Page and Wootters [12] first realized that time is an entanglement phenomenon, which places all the same clock readings in the same story.

More recently, an experiment [13] has shown that a static and entangled state between a clock system and the rest of the universe is perceived as evolving from internal observers testing the correlations between the two subsystems. Any hypothetical external observer would see a static, immutable universe, just as the Wheeler-DeWitt equations predict [14].

Separate instants, labeling the nodes where H ports transform virtual bits into quantum information, do not "flow" together.

The entangled instants, labeling pairs of nodes (n, n-1), where a projector P and a CNOT gate act from node n to node n-1, do "flow" together.

Perhaps it should be further emphasized that here the "flow" of time is something very special to entanglement, because it actually refers to a flow of quantum information and does not have the same meaning as the usual "arrow of time" concept. Furthermore, it must be said that "the temporal intertwining" is only a partial view of what actually happens, that is, the space-time entanglement. In this broader vision, the very concept of causality becomes meaningless. In fact, an "event", which is a pixel B at a time $t_n$, which remains trapped in another event, one pixel A at a time $t_{n-1}$, loses its identity, since the entire entangled state is not separable. Therefore we are not able to establish any causal relationship between the two entangled events. We will come back briefly on this subject in the Conclusions

**4. Looking at the Planck scale**

The new OQGN illustrated in Sect. 3, is a sort of classical-quantum hybrid network in which the nodes are pairs of Hadamard-CNOT ports, plus a projector P. The latter describes the measurement performed by a hypothetical internal observer standing on the node n and observing the node n-1. The resulting scenario is a woven quantum network, in which quantum information is borrowed from the future and stored in the past. This is the logical-quantum computation aspect of the entangled quantum space-time, say, the RHS of the equation ER = EPR [9].

As we will see later, there is also a geometric description, where the extra n connecting links of OQGN play a very important role, as they represent the total number of virtual wormholes that connect two Planckian cells of two consecutive slices. Because the Planckian pixels encode the qubits [3], the wormhole connection is just quantum entanglement. This illustrates the LHS of the ER-EPR conjecture.

The two representations lead to the same conclusion: the space-time itself is entangled at the Planck scale.

In the following, we will illustrate the LHS of the equation ER = EPR in this scenario.

**4.1 Energy of the quantum fluctuations of the metric**

As we have seen in Sect. 2, the total quantum information at time $t_n$ is $I_n = (n+1)^2$, which is the total number of qubits. By the QHP, this number should be equal to the number of pixels of Planck area on the n$^{th}$ slice. However, because of the quantum fluctuation of the vacuum, the effective available information on slice n is: $2n+1$ qubits.

By the conservation principle of quantum information, $\Delta I_n$ should vanish on a slice n. Nevertheless, the missing qubits at time $t_n$ are the virtual states, which contribute to the construction of space-time itself., as it was illustrated in [2].

In Sect. 2, we defined: $\Delta I_{n+1,n} = I_{n+1} - I_n = 2n+3 = 2n'+1 \ (n'=n+1)$, as viewed from the antecedent slice, as an increment of quantum information on the subsequent slice, obtained by the quantum operations on the virtual states.

The same argument can be carried on by defining:

$$\Delta I_{n,n-1} = I_n - I_{n-1} = (n+1)^2 - n^2 = 2n+1 \tag{4.1}$$

as viewed from the subsequent slice.

In fact, $n^2$ is the total number of pixels (qubits) in the past of $t_n$, i.e., from $t_0$ to $t_{n-1}$.

In Eq. (2.3) the expression of the cosmological constant $\Lambda_n$ was given in terms of quantum information $I_n$.

Now, there is a finite variation of the cosmological constant, due to the apparent lack in quantum information:

$$\Delta\Lambda_n = \frac{1}{l_P^2 \Delta I_n}. \tag{4.2}$$

As we have seen, this variation is due to the $n^2$ quantum fluctuations of the vacuum. The same arguments for the apparent non conservation of $I_n$ hold for $\Lambda_n$ as well. As we will see in what follows, $\Delta\Lambda_n$ contributes to the balance of the vacuum energy as the energy of the quantum fluctuations of the metric $\Delta g_n$.

From the GR equations, the vacuum energy is:

$$T_n^{Vac} = -\frac{g_n \Lambda_n}{8\pi} \quad \text{(where we put } G = c = 1\text{)} \tag{4.3}$$

with:

$$\Lambda_n = \delta\Lambda_n + \Delta\Lambda_n \tag{4.4}$$

where:

$$\delta\Lambda_n = -\frac{n^2}{l_P^2 (n+1)^2 (2n+1)} \tag{4.5}$$

is the contribution of the quantum fluctuations of the vacuum (virtual states) to the cosmological constant, and

$$\Delta\Lambda_n = \frac{1}{l_P^2 (2n+1)} \tag{4.6}$$

is the contribution of the quantum fluctuations of the metric to the cosmological constant.

The GR equations in the vacuum can then be rewritten as:

$$T_n^{Vac} = -\frac{g_n}{8\pi}(\delta\Lambda_n + \Delta\Lambda_n). \tag{4.7}$$

The vacuum energy of the $n^2$ virtual states is then:

$$E_n^{\Lambda} = \frac{g_n n^2}{8\pi l_P^2 (n+1)^2 (2n+1)} \tag{4.8}$$

and the gravitational energy $E_n^G$ of the quantum fluctuations of the metric is given by:

$$E_n^G = \frac{g_n}{8\pi l_P^2 (2n+1)}. \tag{4.9}$$

Indeed, in Eqs.(4.8) and (4.9) one should replace $E_n^{\Lambda}$ and $E_n^G$ with the energy densities $\rho_n^{\Lambda} \equiv \frac{E_n^{\Lambda}}{L_n^3}$ and $\rho_n^G \equiv \frac{E_n^G}{L_n^3}$ respectively, where we recall that it is: $L_n = (n+1)l_P$, but the expressions in terms of $E_n^{\Lambda}$ and $E_n^G$ are more suitable to our scopes. However, in the end, we will consider again Eq. (4.9) in terms of the energy density $\rho_n^G$, to explicitly calculate the numerical value of the quantum fluctuations of the metric at the Planck scale.

Note that for $n = 0$ it is $E_0^{\Lambda} = 0$. In fact, there are no virtual states for $n = 0$ because time $t_{-1}$ is unphysical. Then, for $n = 0$, the only contribution to the energy of the vacuum is given by the quantum fluctuations of the metric:

$$E_0^G = \frac{g_0}{8\pi l_P^2} = -T_0^{Vac}. \tag{4.10}$$

Eq. (4.9) can be rewritten as:

$$E_n^G = \gamma \frac{g_n}{A_n^{LQG}} \tag{4.11}$$

where $A_n^{LQG} = A_0^{LQG}(2n+1)$, and $A_0^{LQG} = 8\pi l_P^2 \gamma$ is the minimal area in Loop Quantum Gravity [6], and $\gamma$ is the Immirzi parameter [15].

By differentiating both sides of Eq. (4.9) we get:

$$\Delta E_n^G = \frac{\Delta g_n}{8\pi l_P^2 (2n+1)}. \tag{4.12}$$

## 4.2 The period of the quantum fluctuations of the metric

In this section, we will show that for every slice n, a quantum fluctuation of the metric $\Delta g_n$ occurs on a single pixel of the slice, and the demonstration will be made by absurd.

Let us recall that slice n consists of $2n+1$ Planckian pixels, that is, pixels that have linear dimension equal to the Planck length. It follows that the linear dimension of the total slice n is $(2n+1)l_P$.

Now, let us suppose by absurd that the maximal wavelength $\lambda_n^G$ associated with the quantum fluctuation of the metric $\Delta g_n$ on slice $n$ is:

$$\lambda_n^G = (2n+1)l_P \tag{4.13}$$

corresponding to a frequency:

$$\nu_n^G = \frac{c}{\lambda_n^G} = \frac{c}{(2n+1)l_P}. \tag{4.14}$$

For slice $n=0$, the wavelength $\lambda_0^G$ of $\Delta g_0$ equals the Planck length $l_P$, then $\Delta g_0$ is localized in a single pixel. For $n=1$, $\lambda_1^G = 3l_P$, and then $\Delta g_1$ is "spread" over 3 pixels, on slice $n=2$ it is $\lambda_2^G = 5l_P$ then $\Delta g_2$ is "spread" over 5 pixels, and so on.

The period $T_n^G$ of the quantum fluctuations of the metric $\Delta g_n$ on slice n is given by:

$$T_n^G = \frac{1}{\nu_n^G} = (2n+1) t_P. \tag{4.15}$$

However, the model we are considering was originally built by performing the time-slicing given in Eq. (2.1).

Then, the interval of time between slice n and slice n+1 is $\Delta t = t_P$ for every n and coincides with the period $T_n^G$ given in Eq. (4.15) only for $n=0$. So that, it must hold:

$$T_n^G = t_P \qquad \text{for every n.} \tag{4.16}$$

From Eq. (4.13) it follows:

$$\lambda_n^G = l_P \qquad \text{for every n.} \tag{4.17}$$

This means that for every slice n the quantum fluctuation of the metric takes place on a single pixel of the slice, lasts a Planck time unit, has wavelength equal to the Planck length, and reaches a pixel of the next slice.

An equivalent demonstration can be made by assuming the Wheeler conjecture [8], by which in a region of vacuum of dimension L, the energy of the quantum fluctuations of the metric is of order:

$$E^G \approx \frac{\hbar c}{L}.$$

By taking for granted that in quantum geometrodynamics there is a natural cut-off, that is the Planck length $l_P$, there is a maximal bound for $E^G$, namely, the Planck energy: $E_P \approx \frac{\hbar c}{l_P}$.

In our discrete model, the energy of the quantum fluctuations of the metric has the discrete spectrum:

$$E^G{}_n \approx \frac{\hbar c}{L_n} = \frac{\hbar c}{(n+1)l_P} = \frac{\hbar}{t_n}. \tag{4.18}$$

By differentiating both sides of Eq. (4.18) it follows:
$$\Delta E^G{}_n \approx \frac{\hbar}{\Delta t_n} \, . \tag{4.19}$$
By replacing the above expression of $\Delta E^G{}_n$ in Eq. (4.12) one gets:
$$\Delta g_n \cdot \Delta t_n = 8\pi l_P^2 (2n+1)\hbar \tag{4.20}$$
which looks like an uncertainty relation between the metric and time:
$$\Delta g_n \cdot \Delta t_n \geq \frac{A_0^{LQG}}{\gamma}\hbar \tag{4.21}$$
where we recall that $A_0^{LQG} = 8\pi l_P^2 \gamma$ is the minimal area in LQG.
Eq, (4.21) is saturated for $n=0$:
$$\Delta g_0 \cdot \Delta t_0 = \frac{A_0^{LQG}}{\gamma}\hbar \tag{4.22}$$
where:
$$\Delta g_0 = g_0, \ \Delta t_0 = t_P . \tag{4.23}$$
However, in this model (because the Planck time-foliation) it is:
$$\Delta t_n \equiv t_P \qquad \text{for every n} \tag{4.24}$$
and then it holds:
$$\Delta g_n = \Delta g_0 \equiv g_0 \qquad \text{for every n.} \tag{4.25}$$
The value of the maximal fluctuation of the metric, at the Planck scale, is then:
$$\Delta g_P = \frac{A_0 E_P}{\gamma} \, . \tag{4.26}$$
It is easy to verify that $\Delta g_P$ is dimensionless by recalling that we put since the beginning $G = c = 1$, and that we are considering energy densities.
To explicitly calculate the numerical value of $\Delta g_P$, it is better to rewrite Eq. (4.26) as:
$$\Delta g_P = \frac{8\pi l_P^2 \rho_P G}{c^4} \tag{4.27}$$
where:
$$\rho_P = \frac{E_P}{l_P^3} \quad \text{is the Planck energy density}$$
$$l_P = 1{,}6 \times 10^{-35} \, m$$
$$E_P = 1{,}9 \times 10^9 \, J$$
$$c \approx 3 \times 10^8 \, m/s$$
$$G = 6{,}67 \times 10^{-11} \, \frac{N \cdot m^2}{kg^2} .$$
The numerical value of $\Delta g_P$ is then:
$$\Delta g_P \approx 24{,}3 . \tag{4.28}$$
Note that, by Wheeler, at the Planck scale, the value of the quantum fluctuation of the metric coincides with the vale of the metric itself, namely: $\Delta g_P = g_P$, so that it holds:
$$\Delta g_P = g_P \approx 24{,}3 . \tag{4.29}$$
Actually, the maximal fluctuation of the metric behaves like a tiny wormhole, which connects two pixels: pixel A of slices n and pixel B of slice n+1. In fact, these tiny wormholes do entangle two qubits, $|Q\rangle_A$ and $|Q\rangle_B$, which, by the QHP, are encoded by the two pixels A and B on slice n and on slice n+1 respectively. This will be illustrated in more details in the next sections.

**5. Wormholes and space-time entanglement**
In this section, we will illustrate the mechanism of space-time entanglement in terms of virtual mini wormholes. We will show that such a mechanism is equivalent to the one given in terms of an OP, which was illustrated in Sect. 3. Also, we formulate a new version of the holographic principle, adapted to entangled space-time. Finally, we show that there is no information loss inside the entangling wormholes.

**5.1 Observational photon or wormhole?**
In Sect. 3, we considered a fictitious internal observer who, standing on slice n, and observing slice n-1, used an OP of energy $E_n^{OP} = h\nu_n^{OP}$, where $\nu_n^{OP}$ is the frequency of the photon.
The observation will disturb the metric $g_n$ of the empty space on slice n, causing the fluctuation $\Delta g_n$ of the metric.
The energy $E_n^{OP}$ needed for the observation, is given at the expenses of the vacuum energy. The latter is given by the GR equations in the vacuum, namely:

$$T_n^{Vac} = -\frac{g_n \Lambda_n}{8\pi} \tag{5.1}$$

where in this case $\Lambda_n$ is the "full" cosmological constant $\Lambda_n = \frac{1}{l_P^2 I_n}$ (with $I_n = (n+1)^2$).

By taking into account the contribution of the photon energy, Eq. (5.1) becomes:

$$T_n^{Vac} = -\frac{g_n \Lambda'_n}{8\pi} + E_n^{OP} . \tag{5.2}$$

By differentiating both sides of Eq. (5.2), we get:

$$\frac{\Delta g_n \Lambda'_n}{8\pi} = \Delta E_n^{OP} \tag{5.3}$$

and by the time-energy uncertainty relation it follows, from Eq. (5.3):

$$\Delta g_n \cdot \Delta t_n \geq h \frac{8\pi}{\Lambda'_n} \tag{5.4}$$

that is, a kind of time-metric uncertainty relation, which illustrates how the OP induces an uncertainty $\Delta g_n$ in the metric.
Notice that Eq. (5.3) coincides with Eq. (4.12) for:

$$\Lambda'_n = \frac{1}{l_P^2 (2n+1)} . \tag{5.5}$$

Then, using an OP with wavelength $\lambda_n^{OP} = \frac{1}{\nu_n^{OP}}$ on slice n, which disturbs the metric $g_n$ leading to an uncertainty $\Delta g_n$, is equivalent to associate a gravitational wavelength:

$$\lambda_n^G = \lambda_n^{OP} \tag{5.6}$$

to pre-existing quantum fluctuations of the metric $\Delta g_n$ on slice n.
We wish to make the following remarks.
We assumed that the discrete wavelength $\lambda_n^G$ of the quantum fluctuation of the metric is quantized in units of the Planck length $l_P$.
From Eq. (4.17) it follows that, at the Planck scale $(n=0)$ the quantum gravitational wavelength is equal to the Planck length:

$$\lambda_0^G = l_P \ . \tag{5.7}$$

From Eq. (5.6) it follows that it exists a minimal wavelength for the OP as well, which is equal to the Planck length:

$$\lambda_0^{OP} = l_P \ . \tag{5.8}$$

Now, let us consider a Planck particle, that is a micro black hole, whose Compton wavelength $\lambda_C^{Planck}$ is equal to its Schwarzschild radius $r_S^{Planck}$, and both are of the order of the Planck length. From Eq. (5.6) one might identify the OP at the Planck scale with the Planck particle, as it holds:

$$\lambda_0^G = \lambda_C^{Planck} = r_S^{Planck} \approx l_P \tag{5.9}$$

Then, such an OP at the Planck scale will materialize in a Planck particle, or, in other words, it will be swallowed by a micro black hole. The geometrical analogue of this process will be described in details in the next sub-sections. As we will see, at the Planck scale the quantum fluctuations of the metric are distorted, to the point of becoming mini wormholes connecting two Planckian black holes on two consecutive slices.

The OP-interpretation takes into account quantum information rather than geometry. On the contrary the Gravitational (G)-interpretation takes more into account geometry than quantum information.

However, the two interpretations lead to the same results and to the same conclusion: space-time itself is entangled at the Planck scale.

## 5.2 Virtual wormholes connecting spatial slices

We showed that in this model, the period $T_n^G$ of the quantum fluctuations of the metric is equal to the Planck time, which corresponds to the maximal quantum fluctuation of the metric on a pixel of each slice n.

In fact, by recalling that the slices n and n-1 are at times $t_P(n+1)$ and $nt_P$ respectively, it becomes clear that the maximal fluctuations of the metric are virtual wormholes connecting two pixels of minimal area, one on slice n, and the other on slice n-1.

Moreover, one should recall that, by the QHP, each pixel of a slice, encodes one qubit, and this can be visualized like two spin-1/2 punctures one from the top of the slice, and the other from the bottom. However, the puncture from the bottom of slice n, and the one from the top of slice n-1 are inside the bulk of the wormhole, and then they do not contribute to the area of the pixel, which then results to be the same minimal area of LQG: $A^{LQG} = 8\pi l_P^2 \gamma$ found in the case of one single puncture of spin ½.

The wormhole mouth on slice n (as well as the one on slice n-1) can be viewed as a Planckian black hole, and this allows to use the same value of $\gamma$ obtained in [16] through considerations on black holes entropy in the context of LQG, namely: $\gamma = \dfrac{\ln 2}{\pi\sqrt{3}}$.

Then, in this model, a virtual wormhole connects a single pixel of slice n with a single pixel of slice n-1, having the minimal area of LQG.

## 5.3 The holographic principle for entangled space-time

In the OQGN scenario, the entanglement of qubits was realized by introducing quantum logical gates as the Hadamard and the CNOT at each node. The resulting network is, however, just the logical description of the actual entanglement mechanism produced by the virtual wormholes in the quantum cosmological model.

This sort of duality between the logical and the physical-geometrical descriptions arises from the QHP, and is in agreement with the ER=EPR conjecture.

Then, we formulate the holographic principle for entangled pace-time:

i) Every pixel of Planck area is punctured simultaneously by a spin - ½ from the top and a spin ½ from the bottom (or vice versa) that is, by the QHP, it encodes one qubit.
ii) Entanglement of two qubits requires the identification of the bottom of the surface of one pixel of slice n+1 with the top of the surface of another pixel of slice n.
iii) Such identification is realized by virtual wormholes connecting the two slices, and having two mouths, each of Planck area.
iv) The total number of wormholes at time $t_n$ is $n$ (counting all wormholes, since $t_0$).

See **Fig. 3**

### 5.4 No quantum information loss

Let us recall that each slice n consists of $2n+1$ pixels of Planck area, each one encoding one qubit. One pixel of slice n is connected by a wormhole with one pixels of slice $n-1$, and a second pixel of slice n is connected by another wormhole with a pixel of slice n+1.

An observer on slice n, at time $t_n$, will see 2n pixels which are not connected (separate outgoing links in the OQGN). At time $t_{n+1}$ the observer is on slice $n+1$, and realizes that one formerly unconnected pixel of slice n is now connected with one pixel of slice $n+1$. In summary, the observer on slice n+1 will see only 2n-1(separate) pixels of slice n. There is an apparent information loss of two qubits, but indeed such missing information became spacetime entanglement entropy.
A pair of pixels, for example one on slice n and the other on slice n-1, which are connected by a mini wormhole, are maximally entangled bipartite states (Bell states).
Let us consider for example the Bell state:

$$\frac{1}{\sqrt{2}}\left(|0\rangle_A|0\rangle_B + |1\rangle_A|1\rangle_B\right) \tag{5.10}$$

where the qubit A was a pixel on slice n and the qubit B was a pixel on slice n-1. Now they are maximally entangled trough a mini wormhole to form the Bell state in Eq.(5.10).
The global state is pure, that is the entanglement entropy is zero:

$$H(AB) = 0 \ . \tag{5.11}$$

However, the reduced states A and B are both fully mixed:

$$H(A) = H(B) = 1 \tag{5.12}$$

and the mutual information $I(A:B)$ is maximal:

$$I(A:B) = H(A) + H(B) - H(AB) = 2 \ . \tag{5.13}$$

From Eq. (5.10) it is clear that the pixel A on slice n is not anymore a pure state. In fact, one unit of quantum information has been used to maximally entangle the pixel A with the pixel B on slice n-1. Because of that an observer on slice n will see an apparent loss of one unit of quantum information.
The identification, due to entanglement, of the bottom side of a pixel of slice n with the upper side of a pixel of slice n-1, shows that such an entangled pixel is punctured in total by four spin ½ punctures.
We argue then that the micro wormhole connecting two pixels of LQG minimal area is the geometrical analogue of a spin-2 massless particle, namely a graviton.
It should be stressed, however, that such a structure is not a particle, but a maximally entangled bipartite geometrical-quantum-computational structure, with spin 2.

### 6. Conclusions

The results obtained in this work are listed below.
First of all, the quantum space-time itself is entangled at the Planck scale.
It could be interesting to deepen this result more deeply, through the logic of quantum information. There is a no-go theorem of quantum information logic, the "No self-entanglement theorem" [17], which prohibits a single entity that encodes quantum information from becoming entangled with itself. This theorem is in fact a consequence of the no-erase (no-cloning) theorem [18]. Although at first sight this theorem may seem obvious, in reality it is not, on the contrary its consequences are notable, especially in the case of space-time entanglement.

In fact, if we ask ourselves why a pixel of a given slice does not intertwine with another pixel of the same slice, but only with a pixel of the next slice, we realize the importance of this theorem. Actually, it seems that all the pixels of the same slice behave like a single entity: "slice n with 2n+1 pixels". Thus, the slice can not remain entangled with itself. A similar topic, although in a different context, has been carried out in [19].

In other words, entanglement can not occur in a precise instant $t_n$, where there is slice n, but only during a very small time interval $\Delta t = t_P$, i.e., between slice n and slice n-1. An illustrative case is that of slice n=0, that has a single pixel, which is entangled with one of the three pixels of slice n=1. Even if slice n = 0 is a singlet, this process also continues for all the larger slices.

This tells us that space alone can not be entangled, but that space-time can, and that time plays a very important role in this.

Space-time entanglement can be described in two different ways. The first is strictly quantum computational, the latter is more geometric. This equivalence of the two modalities, which in fact lead to the same results and conclusions, provides formal support to the ER = EPR conjecture. Furthermore, we have found that in this model there is the geometric-computational analogue of the graviton.

These results and conclusions were obtained using different techniques, principles and theories, including quantum networks, QHP, Wheeler quantum foam, LQG and CLQG.

Finally, we wish to make the following two observations.

First, our results seem to indicate that the interpretation of Many Worlds [20] and the Objective Reduction (OR) [21] should both be excluded in the presence of space-time entanglement (work in progress).

Secondly, within the framework of an entangled space-time, the concept of causality becomes meaningless. This was formally illustrated in a previous work [22], in which it was shown that a causal set can not accommodate entangled events, as they are cycles. Thus, quantum space-time theories based on causal sets [23] are limited in the sense that they can not include space-time entanglement.

As we have already said in the Sect. 3.4, it is not really a "flow" of time, which arises in the presence of space-time entanglement, but a flow of quantum information from the future to the past. This allows us to think of the past and the future as two non-separable moments, and in this sense time "flows", even if it would be better to say "entangles". A relationship between time and quantum information has also been discussed, in the context of causal networks, in [24].

Without entanglement, the quantum space-time structure, like that of the original QGN, resembles a causal set with a discrete internal arrow of time, a partial order among the nodes, but without a flow of quantum information. And without the latter, past and future are separate events, which do not share any information. Thus, the usual concept of causality should be revisited in some way in order to include entanglement and become in accord with the process of quantum memory storage in the past.

**Acknowledgements**

I am very grateful to Giuseppe Vitiello, Mario Rasetti and Gianfranco Minati for useful advices and comments. I also wish to thank Eliano Pessa for interesting discussions in the early stages of this work. Acknowledgements are due to Aura Rampazzo, CG illustrator, for the graphs and figures.

**Fig. 1**
**The QGN**
The outgoing links are qubits, the connecting links are virtual states of the vacuum, the nodes are Hadamard gates.

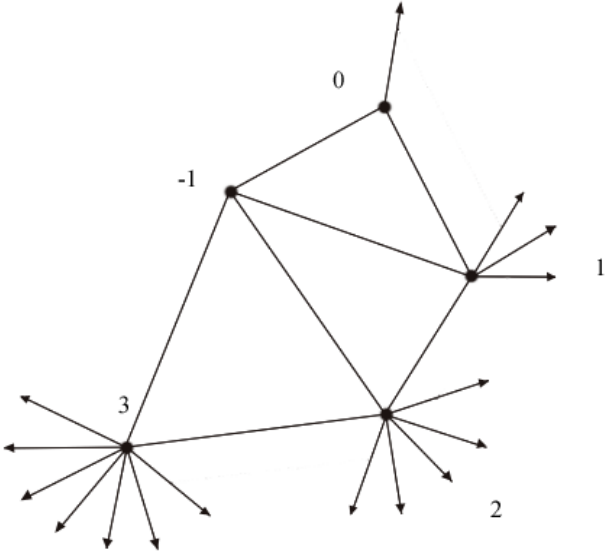

**Fig. 2**
**The OQGN**
The connecting links: straight lines are virtual states of the vacuum, wavy lines are mini wormholes.
Nodes are pairs of quantum gates (**H**, **CNOT**).
The outgoing links are qubits: pairs of thin arrows are entangled states, thick arrows are not entangled states.
**Notice**: the thick arrows sector of the OQGN is the same as the QGN in Fig.1, one time step earlier.

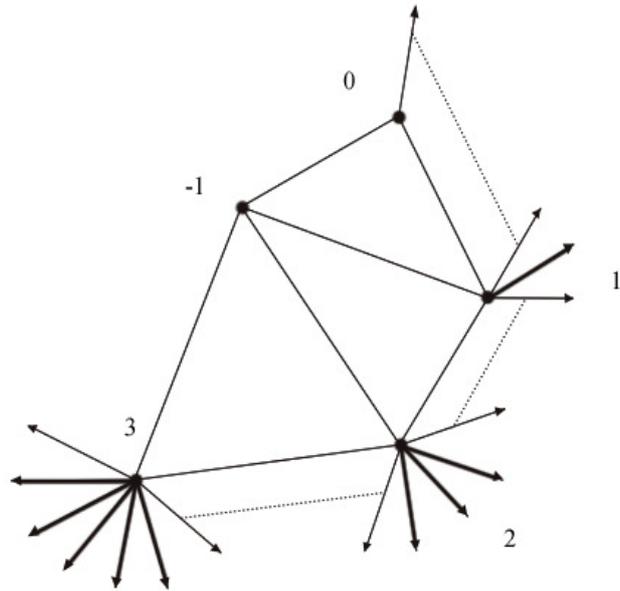

**Fig. 3 a**
**Wormholes connecting two consequent slices**
Two pixels, one of slice n=0 and the other of slice n=1, maximally entangled by a wormhole, in the Bell state: $\frac{1}{\sqrt{2}}\left(|0\rangle_A \otimes |0\rangle_B + |1\rangle_A \otimes |1\rangle_B\right)$

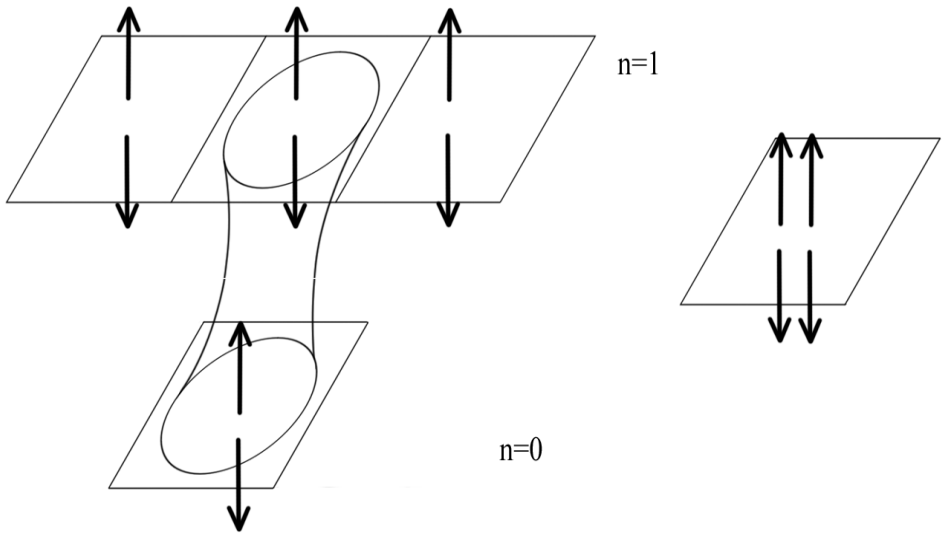

**Fig. 3 b**
Two pixels, one of slice n=0 and the other of slice n=1, maximally entangled by a wormhole, in the Bell state: $\frac{1}{\sqrt{2}}\left(|0\rangle_A \otimes |1\rangle_B + |1\rangle_A \otimes |0\rangle_B\right)$

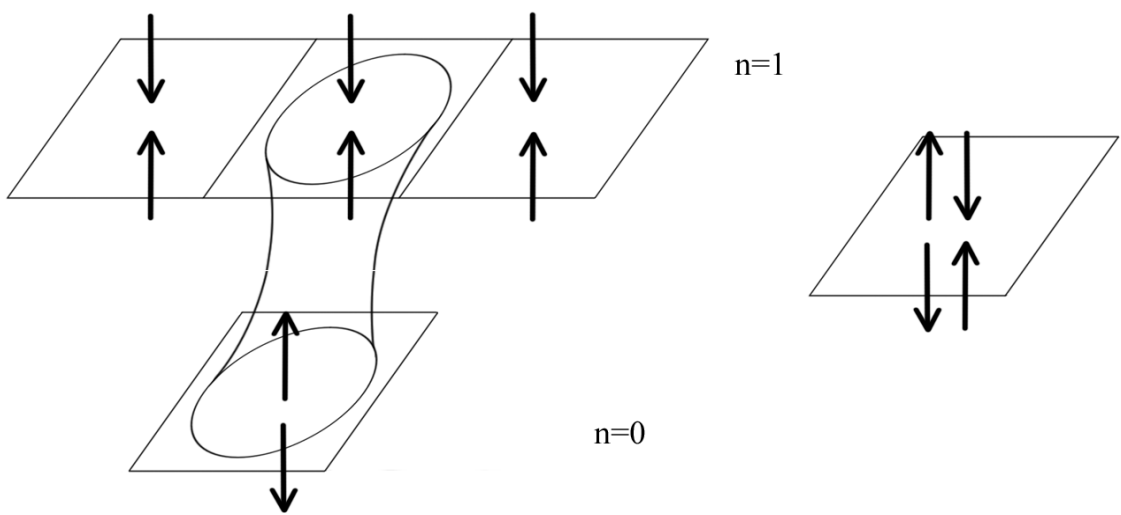